\documentclass[aps,pra, superscriptaddress,twocolumn]{revtex4-1}
\usepackage{graphicx}
\usepackage{epsfig}
\usepackage{amsmath}
\usepackage{amssymb}
\usepackage{amsfonts}
\usepackage{mathrsfs}
\usepackage{theorem}
\usepackage{bm}
\usepackage{url}
\usepackage[T1]{fontenc}
\usepackage{csquotes}
\MakeOuterQuote{"}
\usepackage{fancyhdr}
\usepackage[english]{babel}

\usepackage{dcolumn}
\usepackage{setspace}
\usepackage{color}

\def\vec#1{\bm{#1}} 

\newcommand{\tr}{\operatorname{tr}}

\newcommand{\diag}{\operatorname{diag}}

\newcommand{\rmE}{\mathrm{E}}
\newcommand{\rmd}{\mathrm{d}}

\newcommand{\be}{\begin{equation}}
\newcommand{\ee}{\end{equation}}
\newcommand{\ba}{\begin{align}}
\newcommand{\ea}{\end{align}}

\def\<{\langle}  
\def\>{\rangle}  

\def\eqref#1{\textup{(\ref{#1})}}  
\newcommand{\eref}[1]{Eq.~\textup{(\ref{#1})}}

\newcommand{\esref}[1]{Eqs.~\textup{(\ref{#1})}}

\newcommand{\fref}[1]{Fig.~\ref{#1}}
\newcommand{\Fref}[1]{Figure~\ref{#1}}

\newcommand{\tref}[1]{Table~\ref{#1}}

\newcommand{\cref}[1]{Conjecture~\ref{#1}}
\newcommand{\Cref}[1]{Conjecture~\ref{#1}}

\newcommand{\rcite}[1]{Ref.~\cite{#1}}
\newcommand{\rscite}[1]{Refs.~\cite{#1}}

\begin{document}


\title{Achieving quantum precision limit in adaptive qubit state tomography}
\author{Zhibo Hou}
\affiliation{Key Laboratory of Quantum Information,University of Science and Technology of China, CAS, Hefei 230026, P. R. China}
\affiliation{Synergetic Innovation Center of Quantum Information and Quantum Physics, University of Science and Technology of China, Hefei 230026, P. R. China}
\author{Huangjun Zhu}
\thanks{hzhu1@uni-koeln.de} 
\affiliation{Perimeter Institute for Theoretical Physics, Waterloo, On N2L 2Y5, Canada}
\affiliation{Institute for Theoretical Physics, University of Cologne, Cologne
 50937, Germany}
\author{Guo-Yong Xiang}
\thanks{gyxiang@ustc.edu.cn}
\affiliation{Key Laboratory of Quantum Information,University of Science and Technology of China, CAS, Hefei 230026, P. R. China}
\affiliation{Synergetic Innovation Center of Quantum Information and Quantum Physics, University of Science and Technology of China, Hefei 230026, P. R. China}
\author{Chuan-Feng Li}
\affiliation{Key Laboratory of Quantum Information,University of Science and Technology of China, CAS, Hefei 230026, P. R. China}
\affiliation{Synergetic Innovation Center of Quantum Information and Quantum Physics, University of Science and Technology of China, Hefei 230026, P. R. China}
\author{Guang-Can Guo}
\affiliation{Key Laboratory of Quantum Information,University of Science and Technology of China, CAS, Hefei 230026, P. R. China}
\affiliation{Synergetic Innovation Center of Quantum Information and Quantum Physics, University of Science and Technology of China, Hefei 230026, P. R. China}





\date{\today}
\begin{abstract}
\noindent The precision limit in quantum state tomography is of great interest not only to practical applications but also to foundational studies. However, little is known about this subject in the multiparameter setting even theoretically due to the subtle information tradeoff among incompatible observables. In the case of a qubit, the theoretic precision limit was determined by Hayashi as well as Gill and Massar, but attaining the precision limit in experiments  has remained a challenging task. Here we report the first experiment which achieves this  precision limit  in adaptive quantum state tomography on optical polarization qubits. The two-step adaptive strategy employed in our experiment is very easy to implement in practice. Yet it is surprisingly  powerful in  optimizing most figures of merit of practical interest. Our study may have significant implications for multiparameter quantum estimation problems, such as quantum metrology. Meanwhile, it may promote our understanding about the complementarity principle and uncertainty relations from the information theoretic perspective.

\end{abstract}

\maketitle

\section{INTRODUCTION}

\noindent Quantum state tomography is a procedure for inferring the state of a quantum system from quantum measurements and data processing \cite{PariR04,Haya05book,LvovR09}. It is
a primitive of various quantum information processing tasks, such as quantum computation, communication, cryptography, and metrology \cite{GiovLM04, GiovLM06, GiovLM11, XianHBW11, CrowDBW12}. In sharp contrast with the classical world, any measurement on a generic quantum system necessarily induces a disturbance, limiting further attempts to extract information from the system. Therefore, many identically prepared systems are usually required for reliable state determination. Conversely, the precision limit in quantum state tomography offers a perfect window for understanding the distinction between quantum physics and classical physics \cite{Zhu12the,BuscHOW14,Zhu15IC, ChenZXT14}.

Recently, great efforts have been directed to improving the tomographic efficiency given limited quantum resources \cite{OkamIOY12,MahlRDF13}. For example, adaptive measurements have been realized in experiments, which may improve the scaling of the infidelity in certain scenarios~\cite{MahlRDF13,KravSRH13}. However, most studies have been tailored to deal with specific figures of merit under special settings, such as pure state or single parameter models, which admit no easy generalization to the more challenging and exciting multiparameter estimation problems with general figures of merit. In particular, the tomographic precision limit  in the multiparameter setting is still poorly understood; experimental studies are especially rare. To fill this gap, in this work we report the first experiment that achieves  the quantum  precision limit in adaptive quantum state tomography on optical polarization qubits.

\section{RESULTS}

\noindent \textbf{Quantum Precision limit}

\noindent In practice, the tomographic precision limit  is determined by experimental settings. As
technology advances, it is ultimately limited by  basic
principles of quantum mechanics. One fundamental limit is known as the quantum Cram\'er-Rao (CR) bound  \cite{Hels67, Hels76book, Hole82book, BrauC94}; see  the appendix.
In the one-parameter setting, this bound can be saturated locally by measuring a suitable  observable, which may depend on the parameter point.
To saturate the bound globally, it is usually necessary to employ adaptive measurements. A simple and effective  choice is known as the two-step adaptive strategy~\cite{Naga89O,GillM00}, whose basic idea can be sketched
as follows. Suppose $N$ copies of the true state are available for tomography. First, we  perform a generic informationally complete measurement on $N_1$ copies (usually $N_1\ll N$ especially when $N$ is large)
of the true state   and compute the maximum likelihood estimator (MLE) \cite{PariR04} according to the measurement statistics. Then we perform the optimal measurement
with respect to the estimator on the remaining $N_2=N-N_1$ copies and compute the MLE
again.

In the multi-parameter setting, however,
the quantum CR bound generally cannot be saturated except when the optimal observables corresponding to  different parameters can be chosen to be compatible.  The existence of incompatible  observables underlies the main distinction between quantum state estimation and classical state estimation and is the main reason why  multiparameter estimation problems are difficult  and poorly understood. Up to now, the optimal solutions have been found only for  a few special cases~\cite{Hole82book,Haya05book}, of which the qubit model is the most prominent~\cite{Haya97,GillM00,Zhu12the}.

To devise a good measurement scheme in the multiparameter setting, it is indispensable  to take into account the subtle information trade-off among incompatible observables.
A vivid manifestation  of such tradeoff is the wave-particle duality relation   between fringe visibility $\mathcal{V}$ and path distinguishability $\mathcal{D}$ \cite{JaegSV95,Engl96},
\begin{equation}\label{eq:WPDR}
\mathcal{D}^2+\mathcal{V}^2\leq 1.
\end{equation}
This phenomenon is not limited to the double-slit experiment but presents itself whenever we are trying to extract information about incompatible observables. It is especially important in understanding multiparameter estimation problems.
Suppose the state of the quantum system is parametrized by a set of parameters denoted collectively by $\theta$, then  such trade-off  can be succinctly summarized by  the following inequality derived by Gill and Massar~\cite{GillM00},
\begin{equation}
\tr\{J^{-1}(\theta) I(\theta)\}\leq d-1,
\end{equation}
which is applicable to any measurement  on a $d$-level system.
Here $I(\theta)$ and $J(\theta)$ are the Fisher and quantum Fisher information matrices, respectively (see the appendix). The Gill--Massar (GM) inequality may be seen as a generalization of the wave-particle duality relation in the language of quantum estimation theory.  To appreciate its significance, it is instructive to point out that the upper bound would be $d^2-1$ if all observables in quantum theory were compatible or, equivalently, if the  quantum CR bound could always be saturated.

The GM inequality  imposes a fundamental   precision limit on  quantum state tomography based on  individual (non-collective)  measurements.
For example, it sets a lower bound for the scaled weighted mean square error
(WMSE) of any unbiased estimator~\cite{GillM00, Zhu12the},
\begin{equation}\label{eq:GMboundWMSE}
\mathcal{E}_{W}^{\mathrm{GM}}=\frac{\bigl(\tr\sqrt{J^{-1/2}WJ^{-1/2}}\,\bigr)^2}{d-1},
\end{equation}
where $W$ is the weighting matrix. Note that the GM bound for the WMSE is $\mathcal{E}_{W}^{\mathrm{GM}}/N$ if the sample size is $N$. In the case of a qubit, the GM bound  agrees with the bound derived by Hayashi~\cite{Haya97} and can  always be saturated locally by mutually unbiased measurements~\cite{GillM00,Zhu12the}
(see the appendix). Recently, the GM inequality and GM bound have been turned into a powerful tool for studying a number of foundational issues  entangled with  incompatible observables \cite{Zhu15IC}, such as the complementarity principle \cite{Bohr28} and uncertainty relations  \cite{Heis27}. The cross fertilization of quantum estimation theory and foundational studies is due to lead to deeper understanding of both subjects \cite{Zhu12the,BuscHOW14,Zhu15IC, ChenZXT14}. Determination of the quantum precision limit in the multiparameter setting is thus of primary interest from both practical and foundational perspectives.

\begin{figure}
  \centering
\center{\includegraphics[scale=0.4]{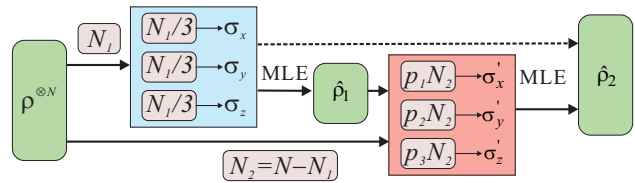}}
  \caption{\label{fig:Adaptive}(color online)  Quantum state tomography with two-step adaptive strategy. The observables $\sigma_x^\prime$,  $\sigma_y^\prime$, $\sigma_z^\prime$ depend on the estimator $\hat{\rho}_1$ obtained in the first step and are related to $\sigma_x$,  $\sigma_y$, $\sigma_z$ via the unitary transformation $U(\hat{\vec{s}}_1)$ described in the main text. The probabilities $p_1$, $p_2$, and $p_3$ depend on both the estimator $\hat{\rho}_1$ and the figure of merit. In the large-$N$ limit, it suffices to use the measurement statistics of step 2 to construct the second MLE. In practice, it is preferable  to employ the measurement statistics of both steps.
   }
\end{figure}

\noindent \textbf{Attaining quantum precision limit  with two-step adaptive strategy}

\noindent Here we report the first experimental verification of the GM bound in adaptive quantum state tomography on optical polarization qubits.
Our tomographic protocol consists of two-step adaptive measurements and MLE, as illustrated in  \fref{fig:Adaptive}. In each step, we need  to  implement only three projective measurements that are mutually unbiased. Despite the simplicity of this approach, it is capable of attaining the quantum precision limit with respect to most figures of merit of interest.
To facilitate applications of our approach, we have  determined  the precision limits and local optimal measurements with respect to a large family of figures of merit in the appendix. For example,   the GM bound for the scaled MSE is given by \cite{HayaM08, Zhu12the}
 \begin{equation}\label{eq:GMBMSEMSBqubit}
\mathcal{E}^{\mathrm{GM}}=\bigl(2+\sqrt{1-s^2}\bigr)^2,
\end{equation}
where $s$ is the length of the Bloch vector of the true state. By contrast, the scaled MSE
achievable by standard tomography using mutually unbiased measurements is given by $3(3-s^2)$ \cite{Zhu12the,Zhu14IOC}.
For concreteness,  the two-step adaptive strategy for minimizing the MSE is sketched as follows:
\begin{enumerate}
\item Measure $\sigma_x, \sigma_y, \sigma_z$ on $N_1/3$ copies of the qubit, respectively, and compute the MLE $\hat{\rho}_1$ based on the measurement statistics. Denote the Bloch vector of $\hat{\rho}_1$ by $\hat{\vec{s}}_1$.

\item Choose a unitary transformation $U(\hat{\vec{s}}_1)$ that rotates $\hat{\vec{s}}_1$ to the $z$ direction and apply this unitary transformation to the remaining $N_2=N-N_1$ copies
of the qubit state.
Measure $\sigma_x, \sigma_y, \sigma_z$ with the following probabilities (see the appendix for more details),
 \begin{equation}\label{eq:GMBMSEoptProb}
p_1=p_2=\frac{1}{2+\sqrt{1-\hat{s}_1^2}}, \quad p_3=\frac{\sqrt{1-\hat{s}_1^2}}{2+\sqrt{1-\hat{s}_1^2}},
\end{equation}
where $\hat{s}_1$ is the length of $\hat{\vec{s}}_1$. Construct the MLE $\hat{\rho}_2$ again based on the measurement statistics.
\end{enumerate}

\begin{figure}[]
  \centering
\center{\includegraphics[scale=0.42]{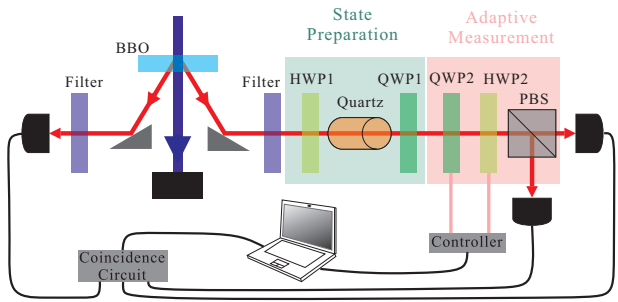}}
  \caption{\label{fig:Setup}(color online) Experimental setup. A pair of horizontally polarized photons are generated via pumping a $\beta$-barium borate (BBO) crystal. One is detected as a trigger and the other is sent through a half-wave plate (HWP), a quarter-wave plate (QWP), and a  770$\lambda$ quartz crystal in between, marked as the state preparation module (green), to change the length and direction of the  Bloch vector. After preparation, the photon polarization is measured by a configuration of QWP2, HWP2, and a polarizing beam splitter (PBS), marked as the adaptive measurement module (pink).}
\end{figure}
\noindent \textbf{Experimental setup}

\noindent The experimental setup is shown in \fref{fig:Setup}. A 2-mm-long BBO crystal cut for type-\uppercase\expandafter{\romannumeral1} phase-matched spontaneous parametric down-conversion (SPDC) process is pumped at 404~nm by a 40-mW V-polarized beam. A pair of H-polarized photons with wave length $\lambda=808$~nm is created via SPDC process. One photon passes through an interference filter whose FWHM is 3 nm, resulting in a coherence length of 270$\lambda$, and is detected by a single photon detector acting as a trigger. The polarization state of the other photon is prepared by HWP1, a 770$\lambda$ quartz crystal, and QWP1. The quartz crystal, whose optic axis is aligned in the horizontal direction, completely decoheres horizontal and vertical polarization components. The rotation angle of HWP1 determines the ratio of H and V polarization components,  thereby together with the quartz crystal changing the purity of the state or the length of the  Bloch vector. QWP1 is used to change the direction of the Bloch vector. QWP2 and HWP2 followed by a PBS are used to perform  arbitrary projective measurements on the qubit. The rotation angles of HWP2 and QWP2 are controlled by a labview program. The coincidence counts are measured by the coincidence circuit and are analyzed by the computer.  Without loss of generality, all  polarization states in our experiments  are prepared such that the Bloch vectors are aligned along  the direction $(0.490,-0.631,0.602)$. This is realized by setting QWP1 at the fixed deviation angle of $19.57^\circ$. The true state is calibrated with about $10^7$ photons.

To achieve  high tomographic precision in experiments,  it is  crucial to reduce the systematic error to a very low level besides adopting the correct adaptive strategy. To this end, we have implemented error-compensation measurements proposed in \rcite{HouZXL15ECM}, wherein multiple nominally equivalent measurement settings are applied to sub-ensembles such that main systematic errors cancel out in the first order. This method is capable of reducing the systematic error from $5\times10^{-5}$ to $8\times10^{-6}$, which is about 100 times smaller than the statistical error when $N=9000$ (see the appendix).

\noindent \textbf{Quantum precision limit with respect to various figures of merit}

\noindent To illustrate the power of the two-step adaptive strategy described above, here we  investigate quantum precision limits with respect to a variety of figures of merit. In the first experiment,  we verify this limit concerning the MSE with  $N_1=3000$ and $N_2=6000$. Here the values of $N_1$ and $N_2$ are determined by numerical simulation to optimize the tomographic precision. The MSE is determined by averaging over 4000 repetitions. As  comparison, we also implement two other tomographic strategies. In the first one, $\sigma_x, \sigma_y, \sigma_z$ are measured on $N/3$ copies of the qubit, respectively, and the MLE is computed as before. This standard tomography is widely used in practice because of its simplicity. In the second one, $\sigma_x, \sigma_y, \sigma_z$ are measured with probabilities $p_1, p_2, p_3$ as specified in \eref{eq:GMBMSEoptProb}  with $\hat{s}_1$ replaced by $s$
after rotating the Bloch vector of each of the $N$ copies of the qubit state to the $z$-axis. This  "known-state tomography" assumes knowledge of the true state in designing the optimal measurement though not in reconstructing the state. It is not feasible in most practical situations, where the state is in fact unknown, but it is useful as a benchmark.

\begin{figure}[]
  \centering
\center{\includegraphics[scale=0.5]{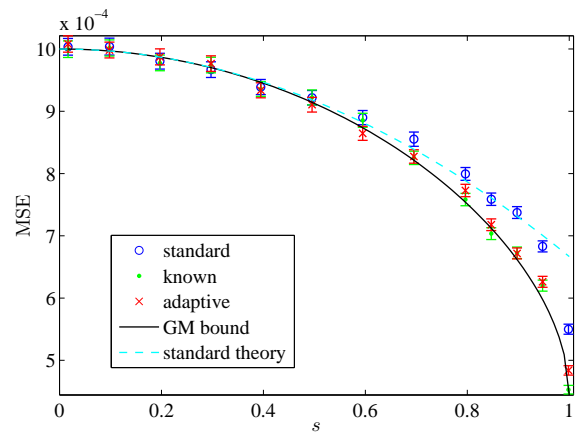}}
\caption{\label{fig:MSE}(color online) Precision limit with respect to the MSE. Experimental results of standard, adaptive, and known-state tomography are shown together with the theoretical MSE of the standard tomography  and the GM bound. Here $s$ is the length of the Bloch vector. In the experiment,   $N=9000$ and $N_1=3000$; each data point averages over 4000 repetitions and the error bar denotes the standard deviation of the mean.  The MSE of the standard tomography is lower than the theoretical value when the purity of the state is sufficiently high (depending on $N$) because the estimator is biased due to the influence of  the boundary of the state space. }
\end{figure}

\Fref{fig:MSE} shows the MSEs associated with the three measurement strategies mentioned above along with the
theoretical MSE of the standard tomography  and the GM bound. The experimental data agree very well with the theoretical prediction.  In contrast with the adaptive strategy proposed in \rcite{MahlRDF13}, which offers no advantage over standard tomography with respect to the MSE, our two-step adaptive strategy is significantly more efficient than standard tomography. What is remarkable,  the MSE achievable by the two-step adaptive scheme is quite  close to that of known-state tomography and  saturates the GM bound approximately.

\begin{figure}[]
  \centering
\center{\includegraphics[scale=0.5]{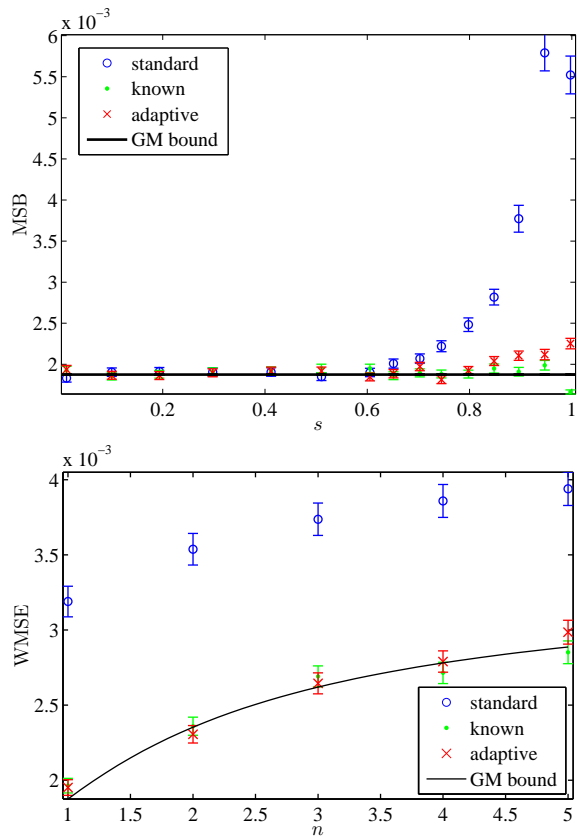}}
\caption{\label{fig:MSB}(color online) Precision limit with respect to the MSB and WMSEs. In the experiment, $N=1200$ and $N_1=300$; each data point is an average over 1000 repetitions and the error bar denotes the standard deviation of the mean.
(upper plot) MSBs of standard, adaptive, and known-state tomography  together with the  GM bound. The MSB of the known-state tomography is slightly smaller than the GM bound when the purity of the true state is sufficiently high (depending on $N$) because the estimator is  biased due to the influence of the boundary of the state space. (lower plot) WMSEs with respect to the family of monotone Riemannian metrics determined by \eref{eq:MonotoneMetricfn}
 for a state with  $s=0.9$.     }
\end{figure}

Next, we investigate the  quantum precision limit  with respect to the mean square Bures distance (MSB).
Incidentally, the GM bound for the MSB is equal to $9/(4N)$ (see  the appendix), which is independent of the unknown qubit state. The two-step adaptive strategy for minimizing the MSB  is quite similar to that for minimizing the MSE except that in the second step $\sigma_x, \sigma_y, \sigma_z$ are measured with probability $1/3$ each. Similar modification applies to the "known-state tomography". Unlike the previous case, it is not necessary to implement  error-compensation measurements \cite{HouZXL15ECM} because a simple measurement strategy can already achieve a sufficiently high precision.
Here our two-step adaptive strategy is  similar to the one proposed in \rcite{MahlRDF13}, which is tailored to minimize the infidelity. This is not surprising since the infinitesimal square Bures distance is equal to the infidelity. Our study provides further justification of the approach in \rcite{MahlRDF13} from a wider context.

The upper plot of \fref{fig:MSB} shows the MSBs associated with  three measurement strategies  along with the theoretical precision limit set by the GM bound. In  standard tomography, the MSB increases rapidly as the purity of the true state increases.
 In sharp contrast, the MSB achievable by the two-step adaptive scheme is almost independent of the purity and is much smaller than that of  standard tomography, the more so the higher purity of the true state.
Moreover, it  saturates the GM bound approximately, except when the true state is nearly pure. The small gap from the bound is mainly due to the fact that the experiment under investigation is not close enough to the asymptotic regime as the ratio $N_1/N$  is nonnegligible. Meanwhile, the efficiency gap of the standard measurement from the local optimal measurement is significant for states with high purities.

A prominent merit of our approach is its versatility in dealing with various figures of merit as emphasized before.
To further demonstrate this point, we now
turn to  verifying the  precision limit with respect to WMSEs based on  monotone Riemannian metrics \cite{Petz96,PetzS96,BengZ06book}; see  the appendix. For concreteness, we shall focus on the
family of  metrics characterized by
the following equation \begin{equation}\label{eq:MonotoneMetricfn}
\rmd l^2=\frac{1}{4}\biggl(\frac{\rmd s^2}{1-s^2}+\frac{s^2\rmd\Omega^2}{f_n(\frac{1-s}{1+s})(1+s)
}\biggr),
\end{equation}
where $\rmd\Omega^2$ is the metric on the unit sphere and $f_n(t)=[(1+t^{1/n})/2]^n$. The GM bound for the scaled WMSE turns out to be
\begin{equation}\label{eq:GMBfn}
\mathcal{E}_{f_n}^{\mathrm{GM}}=\frac{1}{4}\Bigl( \frac{2}{h}+1\Bigr)^2,\quad h=\Bigl[\frac{(1+s)^{1/n}+(1-s)^{1/n}
}{2} \Bigr]^{n/2}.
 \end{equation}
The  metric reduces to the Bures metric when $n=1$  and the quantum Chernoff metric \cite{AudeCMB07} when $n=2$. Other metrics can also be treated on the same footing. The two-step adaptive strategy is quite similar to the one in the previous case, except that in  the second step  $\sigma_x, \sigma_y, \sigma_z$ should be measured with the  probabilities $p_1=p_2=1/(2+ \hat{h})$ and $p_3=\hat{h}/(2+\hat{h})$, where $\hat{h}$ is defined in the same way as $h$ in \eref{eq:GMBfn} but with $s$ replaced by $\hat{s}_1$. The lower plot of \fref{fig:MSB} shows the experimental results for a true state whose Bloch vector has length $0.9$.  The WMSEs achieved by the two-step adaptive strategy are much smaller than that of standard tomography as in the previous case. Moreover, they agree very well with GM bounds.

\section{DISCUSSION}

\noindent  We have implemented optimal adaptive qubit state tomography of mixed states in the multiparameter setting. The two-step adaptive strategy employed in our experiment is very easy to realize in practice. Yet it is surprisingly powerful: It is applicable to optimizing most figures of merit of practical interest, such as the WMSE based on the Bures metric, quantum Chernoff metric, or any other metric commonly used.
Moreover, it  is capable of  attaining the precision limit set by the GM bound approximately. Our experiment represents a significant step towards optimal quantum state tomography in the multiparameter setting, which may have profound implications for multiparameter quantum metrology. Furthermore, our study exemplifies the subtle connection between the complementarity principle and quantum precision limit, thereby promoting the cross fertilization of quantum estimation theory and foundational studies.

\section{METHODS}

\noindent \textbf{Data collection}

\noindent To stabilize the collecting efficiency, multimode fibers are used to direct photons from free space to single-photon detectors. The whole experimental setup is covered by a black hood to keep away  stray light  from outside;  the multimode fibers are wrapped up by black plastic films to suppress laser scattering.   To balance the data-collection time and random coincidences, the power of the continuous pumping laser is adjusted to 40 mW, and the coincidence window is set at 2 ns. The resulting  coincidence rate is 8000 per second, and the coincidence efficiency is about $8\%$.


\section*{ACKNOWLEDGMENTS}

\noindent The authors would like to thank Geoff Pryde for helpful comments and suggestions. The work at USTC is supported by National Fundamental Research Program (Grants No. 2011CBA00200 and No. 2011CB9211200), National Natural Science Foundation of China (Grants No. 61108009, No. 61222504 and No. 11574291). H.Z. is supported  by Perimeter Institute for Theoretical Physics. Research at Perimeter Institute is supported by the Government of Canada through Industry Canada and by the Province of Ontario through the Ministry of Research and Innovation. H.Z. also acknowledges
financial support from the Excellence Initiative
of the German Federal and State Governments
(ZUK 81) and the DFG.

\section*{CONTRIBUTIONS}

\noindent H.Z. devised and analysed the theoretical approach. G.Y.X. devised and designed the experiment. Z.B.H. built the instruments, performed the experiment and collected the data with assistance from G.Y.X. H.Z., Z.B.H. and G.Y.X. analysed the data and developed the interpretation. Z.B.H. performed numerical simulations of the tomographic protocol with assistance from H.Z., C.F.L. and G.C.G. All authors contributed to the manuscript.

\section*{COMPETING INTERESTS}

\noindent  The authors declare no competing financial interests.

\newpage
\appendix

\section{Quantum Cram\'er--Rao  bound}
Suppose the state $\rho(\theta)$ of a given quantum
system is characterized by a set of  parameters $\theta_1, \theta_2,
\ldots, \theta_g$. To determine the values of these parameters, we
may perform  measurements and construct an estimator
based on the measurement statistics. Once a measurement  with outcomes
$\Pi_\xi$ ($\Pi_\xi$ are positive operators that sum up to the identity)  is chosen,  the probability of obtaining each outcome is determined by the Born rule $p(\xi|\theta)=\tr(\rho\Pi_\xi)$. Given an estimator $\hat{\theta}(\xi)$, the accuracy is usually quantified by the mean-square-error (MSE) matrix, with entries  given by
\begin{equation}
C_{jk}(\theta)=\rmE[
(\hat{\theta}_j(\xi)-\theta_j)(\hat{\theta}_k(\xi)-\theta_k)].
\end{equation}
The famous Cram\'er-Rao (CR) bound in statistical inference states that
the MSE matrix $C(\theta)$ of any unbiased estimator is lower bounded by the
inverse of the Fisher information matrix $I(\theta)$; that is, $C(\theta)\geq I^{-1}(\theta)$, where
\begin{equation}
I_{jk}(\theta)=\rmE\biggl[\biggl(\frac{\partial \ln
p(\xi|\theta)}{\partial \theta_j}\biggr)\biggl(\frac{\partial \ln
p(\xi|\theta)}{\partial \theta_k}\biggr)\biggr].
\end{equation}
Meanwhile, $I(\theta)$ is upper bounded by the \emph{quantum Fisher information matrix} $J(\theta)$ with
\begin{equation}\label{sym:QFISLD}
J_{jk}(\theta)=\frac{1}{2}\tr\bigl\{\rho(\theta)[L_j(\theta)L_k(\theta)+L_k(\theta)L_j(\theta)]\bigr\},
\end{equation}
where  $L_j(\theta)$ is the Hermitian
operator that satisfies  the equation
\begin{equation}
\frac{\rmd \rho(\theta)}{\rmd \theta_j}=\frac{1}{2}[\rho(\theta)
L_j(\theta)+L_j(\theta)\rho(\theta)]
\end{equation}
and is known as  the \emph{symmetric logarithmic derivative} (SLD) of
$\rho(\theta)$ with respect to $\theta_j$.
Together with the classical CR bound, the inequality $I(\theta)\leq J(\theta)$ implies
the  quantum CR bound $C(\theta)\geq J^{-1}(\theta)$ \cite{Hels76book, Hole82book, BrauC94}.
 In the one-parameter setting, this  bound can be saturated locally by measuring the observable $L(\theta)$. In the multiparameter setting, however, it generally cannot be saturated except when the $L_j(\theta)$ commute with each other.

\section{Gill--Massar bound for the qubit}
Here we show that the GM bound  for the WMSE in the qubit case can always be saturated, and the local optimal measurement can be realized by a complete set of mutually unbiased measurements \cite{GillM00,Zhu12the,Zhu15IC}. Recall that the GM bound for the scaled WMSE corresponding to the weighting matrix $W$ is given by
\begin{equation}\label{eq:GMboundWMSEsupp}
\mathcal{E}_{W}^{\mathrm{GM}}=\frac{\bigl(\tr\sqrt{J^{-1/2}WJ^{-1/2}}\,\bigr)^2}{d-1}.
\end{equation}
This bound
can be saturated if and only if there exists a  measurement that yields  the
Fisher information matrix
\begin{equation}\label{eq:FisherGM}
I_W=(d-1)J^{1/2}\frac{\sqrt{J^{-1/2}WJ^{-1/2}}}{\tr\sqrt{J^{-1/2}WJ^{-1/2}}}J^{1/2}.
\end{equation}
When $W$ and $J$ commute, \esref{eq:GMboundWMSEsupp}
and \eqref{eq:FisherGM} reduce to
\begin{equation}\label{eq:GMboundWMSE2}
\mathcal{E}_{W}^{\mathrm{GM}}=\frac{\bigl(\tr\sqrt{WJ^{-1}}\,\bigr)^2}{d-1},\quad
I_W=\frac{(d-1)\sqrt{WJ}}{\tr\sqrt{WJ^{-1}}}.
\end{equation}

In the case of a qubit, it is convenient to parametrize the state space by the components of the Bloch vector $\vec{s}=(s_x,s_y,s_z)$. Then the inverse quantum Fisher information matrix takes on the form
$J^{-1}(\vec{s}) =1-\vec{s}\vec{s}$.
Suppose that
$I_W$ in \eref{eq:FisherGM} has eigenvalues $a_1, a_2, a_3$ along with orthonormal eigenvectors $\vec{r}_1, \vec{r}_2,
\vec{r}_3$.  Denote by $s_1, s_2, s_3$ the three components
of the Bloch vector  in this basis; denote by $\vec{\sigma}$ the vector composed of three Pauli matrices $\sigma_x,\sigma_y,\sigma_z$. Then the GM
bound can be saturated by measuring each observable $\sigma_j:=\vec{r}_j\cdot\vec{\sigma}$ with
probability $a_j(1-s_j^2)$. The equality   $\sum_ja_j(1-s_j^2)=\tr(J^{-1} I_W)=1$ \cite{Zhu12the,Zhu15IC} guarantees that the set of probabilities is indeed normalized.
This observation confirms our claim that the optimal measurement scheme   can  be realized using a complete set of mutually unbiased
measurements.

Next, we derive the GM bound when the weighting matrix is the identity or determined by a monotone Riemannian metric \cite{Petz96,PetzS96,BengZ06book}. Without loss of generality, we may assume that  the Bloch vector is aligned with the $z$-axis, so that  $J=\diag(1,1,1/(1-s^2))$, where $s$ is the length of the Bloch vector. If the weighting matrix is  diagonal, say $W=\diag(w_1,w_2,w_3)$, then \eref{eq:GMboundWMSE2} reduces to
\begin{equation}\label{eq:GMBWMSEqubit}
\begin{aligned}
\mathcal{E}_{W}^{\mathrm{GM}}&=\bigl(\sqrt{w_1}+\sqrt{w_2}+\sqrt{w_3(1-s^2)}\bigr)^2,\\
I_W&=\frac{\diag(\sqrt{w_1},\sqrt{w_2},\sqrt{w_3(1-s^2)^{-1}}\,)}{\sqrt{w_1}+\sqrt{w_2}+\sqrt{w_3(1-s^2)}}.
\end{aligned}
\end{equation}
The optimal measurement scheme can be realized by measuring $\sigma_j$ for $j=1,2,3$ ($\sigma_x,\sigma_y,\sigma_z$ in this case) with probability
\begin{equation}\label{eq:OptMeasPro}
p_j=\frac{\sqrt{w_j (1-s^2\delta_{j3} )}}{\sqrt{w_1}+\sqrt{w_2}+\sqrt{w_3(1-s^2)}}.
\end{equation}
The GM bound for the MSE is obtained when $w_1=w_2=w_3=1$, in which case we have (cf. \rscite{HayaM08,Zhu12the})
 \begin{equation}\label{eq:GMBMSEqubit}
\begin{aligned}
\mathcal{E}^{\mathrm{GM}}&=\bigl(2+\sqrt{1-s^2}\bigr)^2,\\
p_1&=p_2=\frac{1}{2+\sqrt{1-s^2}}, \quad p_3=\frac{\sqrt{1-s^2}}{2+\sqrt{1-s^2}}.
\end{aligned}
\end{equation}

Every  monotone Riemannian metric for a qubit has the following form up to scaling,
\begin{equation}\label{eq:MonotoneMetric}
\rmd l^2=\frac{1}{4}\biggl(\frac{\rmd s^2}{1-s^2}+\frac{s^2\rmd\Omega^2}{f(\frac{1-s}{1+s})(1+s) }\biggr),
\end{equation}
where $\rmd\Omega^2$ is the metric on the unit sphere, and  $f$ is a Morozova-Chentsov function \cite{Petz96,PetzS96,BengZ06book}.
If the weighting matrix is determined by this metric, that is,
\begin{equation}
w_1=w_2=\frac{1}{4}[(1+s) f(t)]^{-1},\quad   w_3=\frac{1}{4(1-s^2)},
\end{equation}
where $t=(1-s)/(1+s)$, then
\begin{align}\label{eq:MRMgeneral}
p_1&=p_2=\frac{1}{2+\sqrt{(1+s) f(t)}}, \quad p_3=\frac{\sqrt{(1+s) f(t)}}{2+\sqrt{(1+s) f(t)}}, \nonumber \\
\mathcal{E}_{f}^{\mathrm{GM}}&=\frac{1}{4}\bigl\{2[(1+s)f(t)]^{-1/2}+1\bigr\}^2.
\end{align}
An important family of  Morozova-Chentsov functions has the form \cite{Petz96}
\begin{equation}\label{eq:MCfunction}
f_n(t)=\Bigl(\frac{1+t^{1/n}}{2}\Bigr)^n.
\end{equation}
Accordingly, \eref{eq:MRMgeneral} reduces to
\begin{equation}\label{eq:MRMfn}
\begin{aligned}
p_1&=p_2=\frac{1}{2+ h},\quad p_3=\frac{h }{2+ h },\\
\mathcal{E}_{f_n}^{\mathrm{GM}}&=\frac{1}{4}\Bigl( \frac{2}{h}+1\Bigr)^2,
 \end{aligned}
 \end{equation}
 where
\begin{equation}
h=\Bigl[\frac{(1+s)^{1/n}+(1-s)^{1/n}
}{2} \Bigr]^{n/2}.
 \end{equation}
The  metric turns out to be the Bures metric when $n=1$, in which case $p_1=p_2=p_3=1/3$ and $\mathcal{E}_{\mathrm{B}}^{\mathrm{GM}}=9/4$,  and the quantum Chernoff metric \cite{AudeCMB07} when $n=2$.

%

\section{Systematic error}
In the  ideal scenario, the error  between the calibrated state and the true state vanishes in the large sample limit, so the calibrated state can serve as the true state. In real experiments, however, the error does not vanish due to various experimental imperfections. By the same token, different calibration procedures may  result in different calibrated states. To realize the goal of our experiment, it is crucial that the total systematic error is much smaller than the error due to  statistical fluctuation. This is especially the case in the experiment concerning the MSE, in which the statistical fluctuation is suppressed by  a large sample and large number of repetitions.  In this appendix, we estimate the magnitude of the systematic error and show that it meets the requirement of our experiment.

The systematic errors in our experiment mainly come from six aspects: uncertainties in the phases and rotation angles of  the optical axes  of QWP2 and HWP2, the extinction ratio of the PBS, and  unbalanced collecting efficiencies between the two branches from the PBS. In the analysis of systematic errors, we neglect the statistical fluctuation of measurement results. When the  Pauli operator $\vec{r}\cdot\vec{\sigma}$ with unit vector $\vec{r}$ is measured on $N$ copies of the qubit state $\rho=(1+\vec{s}\cdot\vec{\sigma})/2$, the expected photon counts of the outcomes $\pm1$ are
\begin{equation}\label{photon counts}
  N_\pm=Np_\pm\eta_{\pm},
\end{equation}
where  $p_\pm$ are    probabilities given by  the Born rule and $\eta_{\pm}$ are  collecting efficiencies  determined by  channel losses as well as  coupling and detection
efficiencies  of the two branches.

Consider a PBS with an extinction ratio of $1/\beta$, the  measurement operators corresponding to the two outcomes $\pm1$ actually realized are  $\Pi_\pm=[1\pm(1-2\beta)\vec{r}\cdot\vec{\sigma}]/2$. Accordingly,   $p_\pm=[1\pm(1-2\beta)\vec{s}\cdot\vec{r}]/2$. The unbalance between collecting efficiencies   of the two branches is defined as $\eta=(\eta_+/\eta_-)-1$. To the first  order in  $\eta$, the  frequencies of measurement outcomes are
\begin{equation}\label{frequency}
  f_\pm=\frac{N_\pm}{N_{+}+N_{-}}\approx p_\pm\pm p_+p_{-}\eta.
\end{equation}
The expectation value of $\vec{r}\cdot\vec{\sigma}$ is
\begin{equation}\label{measurement value}
   \hat{m}=f_+-f_-\approx m-2m\beta+2p_+p_{-}\eta,
\end{equation}
where
\begin{equation}\label{expectation value}
  m=\vec{s}\cdot\vec{r}
\end{equation}
is the expectation value if the extinction ratio is infinity and the collecting efficiencies are balanced. In our experiment, the measurement of $\vec{r}\cdot\vec{\sigma}$ is realized  by a configuration of QWP2 and HWP2 as specified by the phases  $\delta_1, \delta_2$ and rotation angles $\theta_1, \theta_2$ of QWP2 and HWP2; see \tref{errorCalculation} for examples.

According to \esref{measurement value} and \eqref{expectation value}, the systematic error of $\hat{m}$ can be estimated as follows,
\begin{equation}
  (\Delta{\hat{m}})^2=\sum_\zeta{(\hat{m}_\zeta)^2(\Delta{\zeta})^2},
\end{equation}
  where $\hat{m}_\zeta$ is the partial derivative of $\hat{m}$ with respect to $\zeta$ with  $\zeta=\beta, \eta, \delta_1, \delta_2, \theta_1, \theta_2$ taken at ideal values of these parameters, that is, $\beta=\eta=0$, $\delta_1=\pi/2$, $\delta_2=\pi$ (the ideal values of $\theta_1$ and $\theta_2$ depend on $\vec{r}$).
 These derivatives are $(\hat{m}_\beta)^2=4m^2$, $(\hat{m}_\eta)^2=4(p_+p_-)^2\leq\frac{1}{4}$; if $\zeta=\delta_1, \delta_2, \theta_1, \theta_2$, then $(\hat{m}_\zeta)^2=(m_\zeta)^2=(\vec{s}\cdot\vec{r}_\zeta)^2\leq|\vec{r}_\zeta|^2$, where $\vec{r}_\zeta=\partial{\vec{r}}/\partial{\zeta}$. Although the dependence of $\vec{r}$ on $\delta_1, \delta_2, \theta_1,\theta_2$ is complicated \cite{HouZXL15ECM}, the magnitudes  of its partial derivatives have simple forms:
 \begin{equation}
 \begin{aligned}
 |\vec{r}_{\delta_1}|^2&=\sin^2(2\theta_1-4\theta_2), \\
 |\vec{r}_{\delta_2}|^2&=\sin^2(2\theta_2), \\
 |\vec{r}_{\theta_1}|^2&=4+4\cos^{2}(2\theta_1-4\theta_2), \\
 |\vec{r}_{\theta_2}|^2&=16.
 \end{aligned}
 \end{equation}
 When $\vec{r}$ is one of the basis vectors, the magnitudes of these partial derivatives are listed in  \tref{errorCalculation}.

 \begin{table}
  \caption{\label{errorCalculation}Rotation angles of QWP2 and HWP2 for realizing the measurement of the Pauli operator $\vec{r}\cdot\sigma$. Also listed are the partial derivatives of $\vec{r}$ with respect to  phases and rotation angles of the two wave plates.}
  \begin{tabular}{c c c c c c }
    \hline\hline
    $\vec{r}$ & $(\theta_1, \theta_2)$ & $|\vec{r}_{\delta_1}|^2$ & $|\vec{r}_{\delta_2}|^2$ & $|\vec{r}_{\theta_1}|^2$ & $|\vec{r}_{\theta_2}|^2$\\
    \hline
    (1, 0, 0) &        (45$^\circ$, 22.5$^\circ$) & 0  & 0.5 & 8  &  16\\
    (0, 1, 0)& (0$^\circ$,  22.5$^\circ$)  & 1 & 0.5 &  4 &  16 \\
    (0, 0, 1) & (0$^\circ$,     0$^\circ$)  & 0  & 0 & 8  &  16 \\
    \hline\hline
  \end{tabular}
  \end{table}

In our calibration process, three Pauli operators $\vec{r}^j\cdot\vec{\sigma}$ are measured, where $\vec{r}^j$ for   $j=1, 2, 3$ form an orthonormal basis, so that the eigenbases of $\vec{r}^j\cdot\vec{\sigma}$ are mutually unbiased.
The calibrated state is determined by the formula
\begin{equation}\label{reconstructed state}
  \hat{\rho}=\frac{1+\sum_{j=1}^{3}{\hat{m}^j\vec{r}^j\cdot\vec{\sigma}}}{2}.
\end{equation}
The systematic error between the calibrated  state and the true  state can be calculated  as follows,
\begin{equation}\label{estimation MSE}
\sum_{\zeta}\sum_{j=1}^{3}(\hat{m}_\zeta^j)^2 (\Delta \zeta)^2=\sum_{\zeta}({M_\zeta})^2(\Delta \zeta)^2,
\end{equation}
where $({M_\zeta})^2=\sum_{j=1}^{3}({\hat{m}_\zeta^j})^2$ and  $\zeta=\beta, \eta, \delta_1, \delta_2, \theta^j_1, \theta^j_2$.

 In order to decrease the systematic error caused by the PBS, we use a beam displacer (BD) acting as a PBS, whose extinction ratio is about 8000. The resulting contribution to the systematic error is $({M_\beta})^2(\Delta\beta)^2\leq6\times10^{-8}$. Notice that
 \begin{equation}
 ({M_\beta})^2=\sum_{j=1}^{3}({\hat{m}_\beta^j})^2=4\sum_{j=1}^{3}({\hat{m}^j})^2=4|\vec{s}|^2\leq4
 \end{equation}
  since $\vec{r}^j$ for $j=1, 2, 3$ are orthonormal.

 The unbalance between the collecting efficiencies  of the two branches is mainly due to the fluctuation caused by  the disturbance of the air flow as well as the vibration and  shift of the laser propagation direction. With a multimode fiber collecting system, this fluctuation can be reduced to below 0.002 within 16 hours, the measurement duration of 4000 repetitions for one quantum state.  This fluctuation results in $(\Delta\eta)^2=4\times10^{-6}$ and $({M_\eta})^2(\Delta\eta)^2\leq3\times10^{-6}$. Notice that $(\hat{m}^j_\eta)^2\leq\frac{1}{4}$ and $({M_\eta})^2=\sum_{j=1}^{3}({\hat{m}_\eta^j})^2\leq\frac{3}{4}$.

The  wave plates provided by our manufacturer have phase uncertainties of about $|\Delta\delta|=1.2^\circ$ from ideal phases, that is, $\delta_1=90^\circ\pm1.2^\circ$ and $\delta_2=180^\circ\pm1.2^\circ$ for QWP2 and HWP2. The phase uncertainties of QWP2 and HWP2 together contribute $[({M_{\delta_1}})^2+({M_{\delta_2}})^2](\Delta\delta)^2\leq8.8\times10^{-4}$ to the systematic error. This systematic error would wash out the advantage of adaptive state tomography over standard tomography when $N=1200$. To solve this problem, we use  calibrated phases of $\delta_1=88.7^\circ\pm0.3^\circ$ and $\delta_2=179.9^\circ\pm0.3^\circ$ provided by the manufacturer to calculate the rotation angles  required to realize  desired measurement settings. Consequently, the systematic error is reduced to $[({M_{\delta_1}})^2+({M_{\delta_2}})^2](\Delta\delta)^2\leq5.5\times10^{-5}$.

The rotation stages used in our experiments have a precision of $0.01^\circ$; the uncertainties in rotation angles $\theta_1$ and $\theta_2$ of QWP2 and HWP2 are mainly determined by the uncertainties in the calibration angles of their optical axes, which are about $\Delta\theta= 0.1^\circ$. The resulting systematic error is  $[(M_{\theta_1})^2+(M_{\theta_2})^2](\Delta\theta)^2\leq 2\times10^{-4}$.

The  total systematic errors due to the six aspects of experimental imperfections mentioned above is no larger than $2.6\times10^{-4}$. Incidentally, the  error between states calibrated by standard tomography and known-state tomography with about $10^{7}$ photons in the experiment is around $5\times10^{-5}$. In our experiments concerning the MSB and WMSEs with respect to a family monotone Riemannian metrics, this systematic is low enough to guarantee that the experimental result agrees well with the theoretical prediction and that the advantage of adaptive state tomography over standard tomography is clearly manifested.

In the experiment concerning the MSE, however, the systematic error is still not small enough since the advantage of adaptive strategy over standard strategy is quite limited and we need to adopt a large sample size and a large number of repetitions to overcome this issue. To solve this problem, we developed and implemented  error-compensation measurements  \cite{HouZXL15ECM}, wherein multiple nominally equivalent measurement settings are applied to sub-ensembles such that main systematic errors cancel out in the first order. This method alleviates  the influences of uncertainties in the phases and the calibration angles of  QWP2 and HWP2. In this way, the total systematic error can be reduced to  $7\times10^{-6}$. Accordingly, the  error between the two calibrated states by standard tomography and known-state tomography with about $10^{7}$ photons  is reduced to  $8\times10^{-6}$ \cite{HouZXL15ECM}.
Now the systematic error is much smaller than the MSE gap between adaptive tomography and standard tomography. Consequently, the experimental result agrees well with the theoretical prediction again.

\bibliographystyle{npjquantuminformation}
\bibliography{all_references}

\end{document}